\documentclass[aps,preprint,floats,prb]{revtex4}
\usepackage{graphics,dcolumn}

\begin{document}

\title{Simulation of Heme using DFT+U: a step
toward accurate spin-state energetics}
\author{Dami\'{a}n A. Scherlis$^{\dag}$}
\author{Matteo Cococcioni$^\S$}
\author{Patrick Sit$^{\ddag}$}
\author{Nicola Marzari$^{\ddag}$}
\affiliation{$^\dag$Departamento de Qu\'{i}mica Inorg\'{a}nica, Anal\'{i}tica
y Qu\'{i}mica F\'{i}sica, Facultad de Ciencias Exactas
y Naturales, Universidad de Buenos Aires, Ciudad Universitaria,
Pab. II, Buenos Aires (C1428EHA) Argentina}
\affiliation{$^\S$Department of Chemical Engineering and Materials Science,
University of Minnesota, Minneapolis MN 55455}
\affiliation{$^\ddag$Department of Materials Science and Engineering,
and Institute for Soldier Nanotechnologies,
Massachusetts Institute of Technology, Cambridge MA 02139}

\begin{abstract}
We investigate the
DFT+U approach as a viable solution to describe
the low-lying states of ligated and unligated iron heme complexes.
Besides their central role in organometallic chemistry, these compounds represent a
paradigmatic case where LDA, GGA, and common hybrid
functionals fail to reproduce the experimental magnetic splittings.
In particular, the imidazole pentacoordinated heme is incorrectly
described as a triplet by all usual DFT flavors. In this study
we show that a U parameter close to 4 eV leads to spin transitions and molecular geometries
in quantitative agreement with experiments, and that
DFT+U represents an appealing tool in the description of iron porphyrin complexes,
at a much reduced cost compared to correlated quantum-chemistry
methods. The possibility of obtaining the U parameter from first-principles is
explored through a self-consistent linear-response
formulation. We find that this approach, which proved to be successful in other iron systems,
produces in this case some overestimation with respect to the optimal values of U.

\end{abstract}

\date{\today}
\maketitle

\section{Introduction}

Enzymatic sites containing transition metals are
among the most relevant biophysical systems currently studied
using first-principles quantum mechanical approaches.
The application of such tools, however, is often severely limited
as a consequence of the inability of conventional electronic
structure methods---such as Hartree-Fock or density-functional theory---to
provide a qualitatively correct description of the spin-state
energetics of the metal center.
Iron porphyrins, which constitute the prosthetic group of the ubiquitous
heme proteins, are a paradigmatic example where the aforementioned
approaches can not be relied upon to predict the ground state multiplicity
of the system.

\begin{figure} \centerline{
\resizebox{4in}{!}{\includegraphics{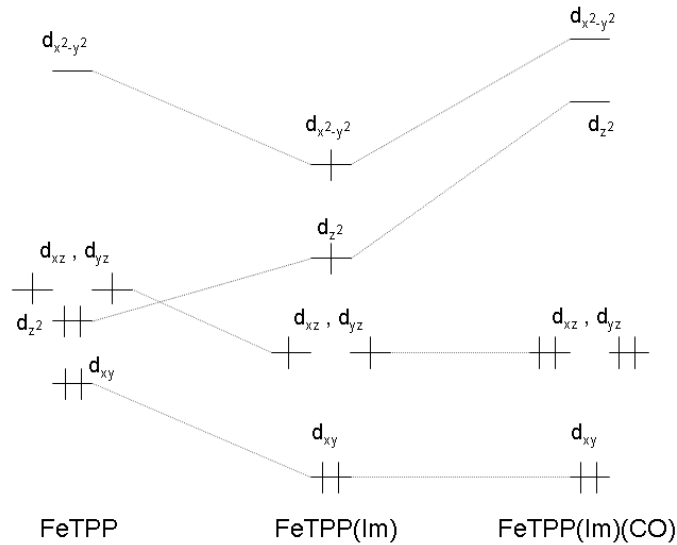}}
}
\caption{Schematic representation of the d-orbitals energy levels for FeTPP. From left to right: free
(four coordinated), ligated to imidazole (five coordinated),
and ligated to imidazole plus CO (six coordinated).}
\label{dorbitals}
\end{figure}

The spin state of iron porphyrins, as much as the spin state of any transition
metal complex, is determined by the coordination symmetry and the nature 
of the ligands. The three lowest accessible spin states (a singlet, a
triplet and a quintuplet if the number of electrons is even, or
a doublet, a quartet and a sextet if it is odd) are conventionally
referred to as low, intermediate, and high-spin.
In unligated porphyrins the metal is coordinated
to four in-plane nitrogen atoms, and experimental studies on model compounds,
namely on Fe(II) tetraphenylporphine (FeTPP), indicate for this coordination
mode a triplet ground state.\cite{t1}$^-$\cite{t5} Additional axial ligands
produce alternative multiplicities: imidazole gives rise to high spin hemes,
while strong ligand-fields as that of diatomic molecules like CO, NO or O$_2$,
favor low spin configurations.\cite{revscheidt} Six coordinated hemes, with
two axial ligands, usually exhibit a low spin state unless the ligand-field
is extremely weak. Fig.~\ref{dorbitals} depicts a schematic view of
the d-states energy levels in three distinctive coordination environments.

Even though first-principles approaches, specifically
Hartree-Fock (HF) and density-functional theory (DFT), greatly contributed
to the interpretation and understanding of the functional aspects of the
active site of heme proteins at the molecular level, attempts to predict the
ground state multiplicity of these systems soon made apparent that an 
accurate description of the electronic structure
might require more sophisticated techniques. This fact can be tracked down to the
spin-transition energies provided by HF and DFT for isolated iron atoms and ions
or different iron compounds, where it has been systematically observed that HF
favors high-spin electronic configurations while DFT exhibits a preference for
low-spin states.\cite{ijqc}$^-$\cite{harvey} Such biases are similarly manifested
in heme complexes: Table I summarizes this trend in five and six
coordinated iron porphines (FeP). 

For the last decade DFT has been the first method of choice to perform 
electronic structure calculations of biological models, and in particular of 
heme systems. In this context, one of the most 
crucial failures of common exchange-correlation functionals has 
been detected in the deoxygenated active site of hemoglobin and 
myoglobin (Table I). The earliest study reporting this flaw is due
to Rovira $et~al.$,\cite{rovira} who obtained for the five coordinated model Fe$\mathbf{^{II}}$P(Im)
(axial ligand: imidazole) a triplet state 6.5 kcal/mol below the quintuplet,
which is the experimental ground state of the system. After this work, a few
others followed which also observed this inversion using B3LYP or different pure
GGA functionals.\cite{ijqc,spiro}
Liao and Scheiner claimed to have found a quintuplet ground
state for this five coordinated compound employing a DFT-GGA 
functional.\cite{scheiner} In their calculations,
however, electronic symmetry constraints were imposed. To the best of our
knowledge, DFT functionals yield for Fe$\mathbf{^{II}}$P(Im) a triplet ground state in
the absence of symmetry constraints. In an effort
to quantify the errors in the DFT estimates of spin transition energies,
Ghosh and Taylor resorted to highly-correlated techniques as CASPT2 and CCSD(T) 
to explore the iron (III) porphyrin chloride.\cite{ghosh} This is another example
of a high spin five coordinated heme complex for which DFT predicts a quartet
favored over the sextet, in this case by around 7 kcal/mol. B3LYP, on the other
hand, finds about the same energy for both spin configurations. The more accurate
approaches CASPT2 and CCSD(T) agree in yielding a sextet ground state, 16 kcal/mol
below the quartet.\cite{ghosh} A latter work by these authors shows the same
low-spin bias in B3LYP for the iron (IV) porphyrin difluoride.\cite{ghosh1} 
It is worth noting here that even CASPT2---employed with the moderate
active spaces currently affordable---has been found fallible in the estimation of
these elusive spin states. Inaccuracies have been reported in the prediction of
the electronic ground states of the isolated iron porphyrin\cite{jcp-choe} and of the 
oxyheme.\cite{jensen}

It is possible to find a rationale for the biases in
DFT and HF, considering the balance between the computed electronic exchange
and correlation energies. In a simplified picture, the (negative) exchange energy
is contributed by like spin electron pairs, while electronic correlation arises
from the interaction between electrons regardless of their spin. A method which
includes the exchange and neglects the correlation, as HF does, will
favor high multiplicities by maximizing
the number of electrons with the same spin. To the contrary, experience shows that
the combination of the exchange and the correlation terms in pure DFT pushes the
balance toward low spin configurations.\cite{harvey} Attempts to improve the spin
state energetics description of density-functionals have mostly been based on
hybrid Hartree-Fock/DFT schemes,\cite{ijqc,deeth,harvey} which combine the exchange
of HF with the exchange and correlation obtained from DFT in proportions 
obeying empirical considerations. This approach, however, has given no universal 
functional capable to provide accurate splittings in every case. In general, those 
functionals offering a good description of the high spin species fail when tried out 
on low spin complexes, and vice versa.\cite{ijqc} Among them, B3LYP is seemingly 
the one with the best average performance up to now, yet exhibiting serious
inaccuracies in the five coordinated models already discussed. 

In the present study, we propose the DFT+U approach as an
alternative to the standard ab-initio techniques for a reliable description
of the low lying states of iron heme complexes.
The LDA+U or GGA+U method (more generally denoted as DFT+U)
was originally designed within the density-functional theory
framework for the treatment of strongly correlated materials.\cite{anisimov1}$^-$\cite{dudarev}
Only very recently researchers have started to apply it to molecular,
or mixed solid-molecular systems,
with extremely promising results.\cite{prl-u}$^-$\cite{ceria2}
This approach corrects the tendency to overhybridize and delocalize electronic
orbitals---ultimately originating in the presence of self-interactions in the
exchange-correlation functionals---by introducing a term that penalizes fractional occupancies.
We note in passing that in our present implementation we explore the possibility of a $U$ that is not a
best-fit parameter, but an intrinsic, ab-initio linear-response property of the system chosen.
However, this approach does not prove totally satisfactory, as in the case of low spin complexes
it leads to values of $U$ lying 1 or 2 eV above the optimal ones.
We show that with the inclusion of a single parameter, DFT+U recovers the correct multiplicities of
the five coordinated models where DFT and hybrid methodologies are in 
disagreement with more elaborated techniques or experimental data.
Moreover, there is no impairment with respect to GGA functionals in those
cases for which DFT displayed the right behavior. Calculations of
ligand exchange thermodynamics, spin transitions, and other properties,
point to GGA+U as an appealing tool to overcome the limitations
entailed by the use of DFT in the description of bioinorganic complexes,
at a computational expense much lower than demanded by highly-correlated
quantum chemistry methods.

\section{Methodology}

\subsection{General framework}
All calculations reported in this work have been performed with the
public domain PWSCF and CP codes included in the Quantum-Espresso
distribution,\cite{Espresso} based on density-functional theory,
periodic-boundary conditions, plane-wave basis sets, and
pseudopotentials to represent the ion-electron interactions.
The PBE exchange-correlation functional\cite{PBE} has been used in
combination with Vanderbilt ultrasoft pseudopotentials,\cite{usp} with the
Kohn-Sham orbitals and charge density expanded in plane
waves up to a kinetic energy cutoff of 25 and 200 Ry respectively.

\subsection{The DFT+U approach}

The present implementation of DFT+U stems from the early contributions by Anisimov and others,
\cite{anisimov1}$^-$\cite{dudarev} who proposed to correct the failures of the LDA
functional in dealing with the strongly localized d or f electrons of transition metal ions.
An on-site correction was thus constructed to account for strong electronic correlations poorly
described within the local-density or generalized-gradient approximations,
and formulated as follows:
\begin{equation}
E_{DFT+U}[\rho({\bf r})]=E_{DFT}[\rho({\bf r})] + 
E_{U}[\{n^{I\sigma}_{mm'}\}]=E_{DFT}[\rho({\bf r})] +
E_{HUB}[\{n^{I\sigma}_{mm'}\}] - E_{DC}[\{n^{I\sigma}\}]
\end{equation}
where $\rho({\bf r})$ is the electronic density, $n^{I\sigma}_{mm'}$
are generalized atomic orbital occupations with spin $\sigma$ associated to the $I$ atom, and
$n^{I\sigma}$ is the sum of the occupations corresponding to all
eigenstates, $\sum_m n^{I\sigma}_{mm'}$.
$E_{DFT}[\rho({\bf r})]$ is the standard LDA or GGA energy functional, and
$E_{HUB}[\{n^{I\sigma}_{mm'}\}]$ represents the ``correct'' on-site correlation
energy. Since $E_{DFT}[\rho({\bf r})]$ already contains an approximate
correlation contribution, a term intended to model such a contribution,
$E_{DC}[\{n^{I\sigma}\}]$, must be subtracted to avoid double counting.

In this work, we resort to the rotationally invariant formulation of DFT+U
introduced by Liechtenstein $et~al.$\cite{anisimov4} and later simplified by
Dudarev and his coworkers,\cite{dudarev} in which the non sphericity of the electronic interactions
and the differences among the interactions in like-spin and unlike-spin
channels are neglected. With these assumptions, the correction to the energy functional
can be written
\begin{equation}
E_{U}[\{n^{I\sigma}_{mm'}\}]=\frac{U}{2} \sum_I \sum_{m,\sigma}[n^{I\sigma}_{mm}-
\sum_{m'}n^{I\sigma}_{mm'}n^{I\sigma}_{m'm}]=
\frac{U}{2} \sum_{I,\sigma} {\bf Tr} [n^{I\sigma} (1-n^{I\sigma})]
\end{equation}
where $U$ is the Hubbard parameter describing on-site correlations.
In principle, different definitions for the occupation matrix are possible, which in
turn will determine different values for $U$. In this case we define
\begin{equation}
n^{I\sigma}_{mm'}=\sum_\nu f_\nu \langle \psi^\sigma_\nu|\phi^I_m \rangle 
\langle \phi^I_{m'}|\psi^\sigma_\nu \rangle
\end{equation}
with $f_\nu$ the weight of the electronic state $\nu$, $\phi^I_m$ the valence
atomic orbital $|lm\rangle$ of atom $I$, and $\psi^\sigma_\nu$ the one electron wavefunction corresponding
to the state $\nu$ with spin $\sigma$. The diagonalization of the occupation matrices
leads to the following expression for the energy correction:
\begin{equation}
E_{U}[\{n^{I\sigma}_{mm'}\}]=\frac{U}{2} \sum_{I,\sigma} \sum_i \lambda^{I\sigma}_i
(1-\lambda^{I\sigma}_i).
\end{equation}
Equation (4) clearly reflects the nature of the correction, which imposes a penalty
(mediated by $U$) for fractional occupations, thus favoring either fully occupied or
empty orbitals ($\lambda \approx$1 and $\lambda \approx$0, respectively).
We note that under this definition, $U$ corresponds to the difference $U-J$ as
utilized by Anisimov and other researchers.\cite{anisimov3}$^-$\cite{dudarev}
For example, the adoption of $U$=4 eV in the present calculations is comparable to
a $U$ of 5 eV in combination with a $J$ of 1 eV in the work of Rollmann.\cite{fedimer}
Whereas in recent applications $U$ is considered a fitting parameter,\cite{fedimer}$^-$\cite{ceria2} 
here we obtain it from the spurious curvature of the DFT energy as a function of the occupations.
As shown by Cococcioni and de Gironcoli,\cite{prb-matteo} the value of $U$ can be 
estimated  as the difference between the screened and bare second derivative 
of the energy with respect to the occupations:
\begin{equation}
U=\frac{\partial^2E_{DFT}}{\partial(n^I)^2}-
\frac{\partial^2E^0_{DFT}}{\partial(n^I)^2}.
\end{equation}
In particular, we are interested in the self-consistent $U$, which we
will call $U_{sc}$, originating from the curvature of the DFT+U ground state.\cite{prl-u}
To compute $U_{sc}$, a few linear response calculations must be performed
at a finite $U_{in}$, each one yielding a corresponding $U_{out}$. It can be shown that
there is a linear dependence between $U_{in}$ and $U_{out}$,
from which $U_{sc}$ can be extrapolated:\cite{prl-u}
\begin{equation}
U_{out}=\frac{\partial^2E_{quad}}{\partial(n^I)^2}=
U_{sc}-\frac{U_{in}}{m}.
\end{equation}
$E_{quad}$ groups all electronic terms within the DFT+U functional
that have quadratic dependence on the occupations, whereas $m$ can be
interpreted as an effective degeneracy of the orbitals whose population
is perturbed. This procedure, which allowed us to attain
an improved description of the multiplet splittings and bonding in
Fe dimers and FeO related species,\cite{prl-u} is the one adopted
here to calculate a self-consistent $U$ parameter for the iron-porphyrin system.
Another criterion has also been explored,
requesting that a linear response calculation at a finite $U$ returns this same
value of $U$ at the output, e.g. $U_{in}=U_{out}$. 
The parameter fulfilling this criterion will be hereafter denoted $U'_{sc}$.
This second criterion is not as appealing as the first one,
since $U_{sc}$ seems the ``right definition'' for self-consistency.

\section{Results and discussion}

In this section DFT+U results are presented on four heme complexes: Fe$\mathbf{^{II}}$P(Im),
Fe$\mathbf{^{III}}$P(Cl), Fe$\mathbf{^{II}}$P(Im)(O$_2$), and Fe$\mathbf{^{II}}$P(CO). 
In the first two cases, DFT calculations fail to predict the high-spin nature of the system.
The other two are examples of low-spin hemes whose
electronic and geometrical properties are, in principle, correctly captured by
standard density-functional simulations. These four case studies were chosen for their
respective relevance in bioinorganic chemistry, and to illustrate the performance
of the DFT+U method on heme models exhibiting a variety of coordination modes
and multiplicities.

\subsection{Five coordinated heme-imidazole complex}

\begin{figure} \centerline{
\resizebox{3in}{!}{\includegraphics{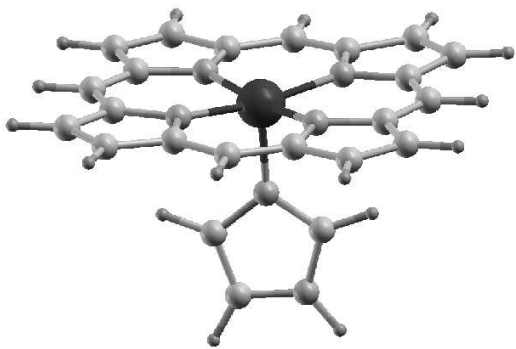}}}
\caption{Structure of the five coordinated Fe$\mathbf{^{II}}$P(Im) complex.}
\label{FePIm}
\end{figure}

The Fe$\mathbf{^{II}}$P(Im) system, depicted in Fig.~\ref{FePIm}, has been the
target of numerous computational studies, inasmuch as it appears as the natural model
to represent the unbound active site of several heme proteins, in particular
hemoglobin and myoglobin.\cite{ijqc,rovira}$^-$\cite{scheiner,jibio}
As mentioned above, the ground state of this compound
has been experimentally characterized as a quintuplet (S=2), whereas
DFT calculations yield a triplet ground state (S=1).
In Table II the energetic separations 
between the low lying spin states resulting from DFT and DFT+U
are compared. According to PBE, the triplet is around 8 kcal/mol
more stable than the quintuplet or the singlet; similar gaps
are obtained with the BP86 exchange-correlation
functional.\cite{rovira} On the other hand,
the transition energy between the triplet and the quintuplet
is reduced to nearly 2 kcal/mol if B3LYP is used.\cite{spiro} As previously noted,
the introduction of the HF exchange in the DFT functional stabilizes
high multiplicity states, but in this case B3LYP is still unable
to provide the right splittings.

\begin{figure} \centerline{
\resizebox{4in}{!}{\rotatebox{-90}{\includegraphics{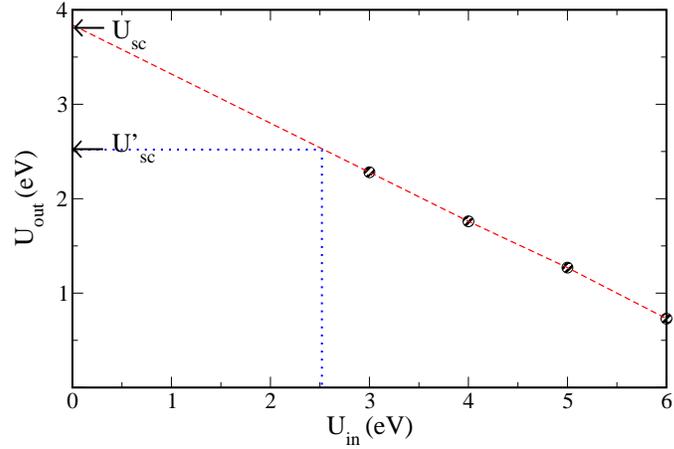}}}}
\caption{Linear response calculation of the $U$ parameter
on the quintuplet state of the Fe$\mathbf{^{II}}$P(Im) complex.}
\label{UvUim}
\end{figure}

\begin{figure} \centerline{
\resizebox{4in}{!}{\rotatebox{-90}{\includegraphics{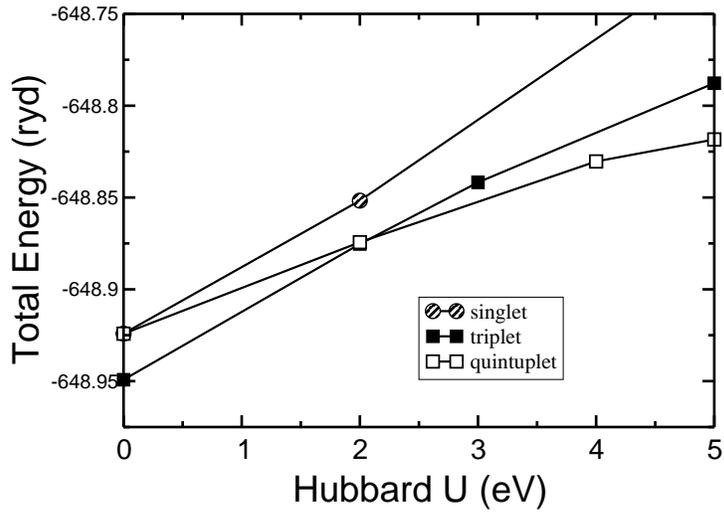}}}}
\caption{Total energy of the low lying spin states of the
Fe$\mathbf{^{II}}$P(Im) complex as a function of the $U$ parameter.}
\label{EvsU}
\end{figure}

The formalism summarized in equation (6)
gives for this system a $U_{sc}$ of 3.9 eV and a $U'_{sc}$ of 2.5 eV (Fig. \ref{UvUim}).
If any of these values are adopted, DFT+U 
restores the experimental ordering of the spin states. The total energies of the low
lying states as a function of $U$ are depicted in Fig.~\ref{EvsU}.
The increase of the $U$ parameter equalizes the triplet and the quintuplet energies,
producing a spin crossover at $U \approx$ 2 eV. At higher values of $U$, the quintuplet
remains the ground state.

\begin{figure} \centerline{
\resizebox{4in}{!}{\rotatebox{-90}{\includegraphics{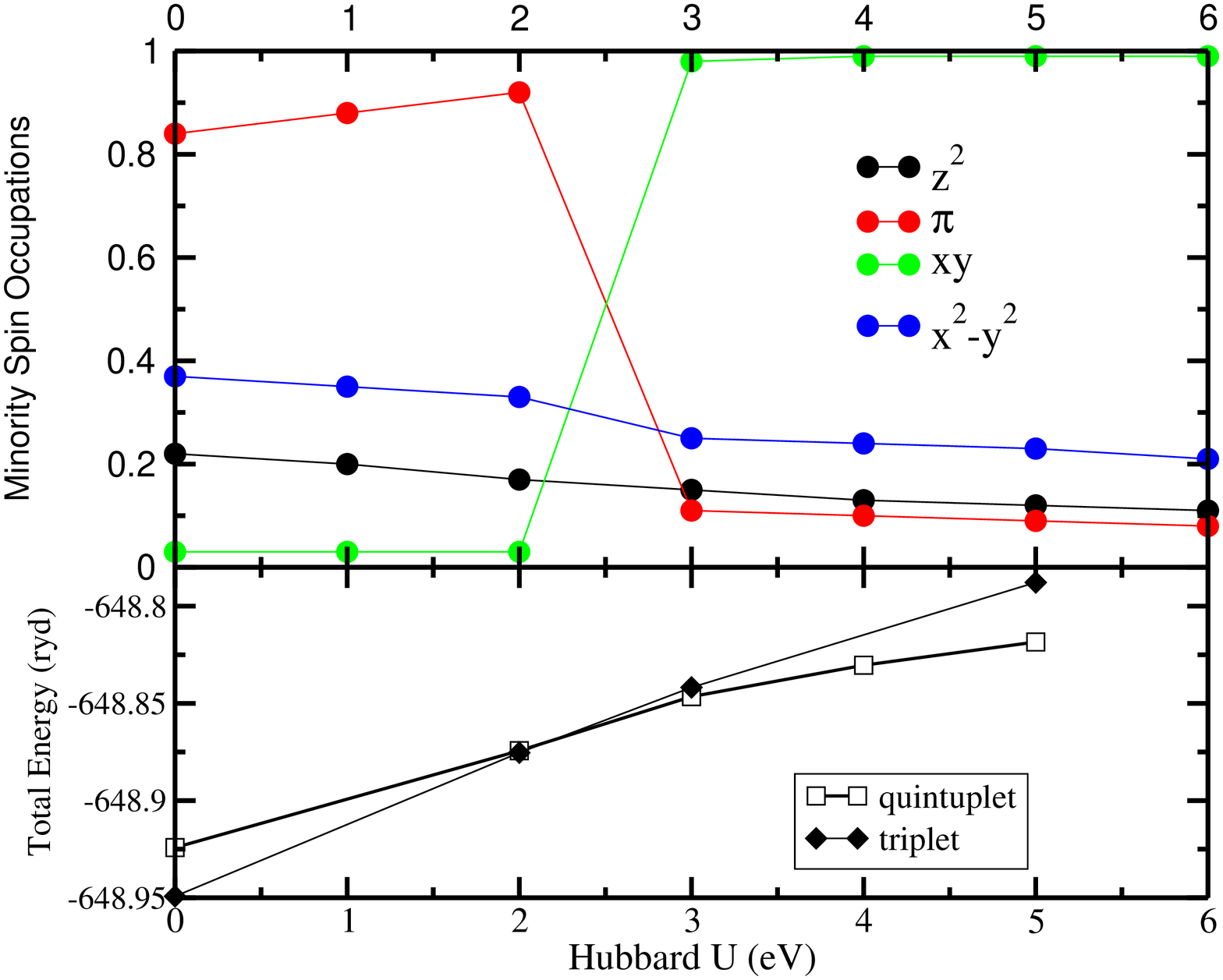}}}}
\caption{Occupations of the minority spin
manifold in the quintuplet state of the Fe$\mathbf{^{II}}$P(Im) complex.
The lower panel shows the total energy of the two lowest accessible spin
states as a function of $U$.}
\label{symm}
\end{figure}

Fig.~\ref{symm} highlights the effect of $U$ on the
electronic symmetry of the d states
in the heme porphyrin. Spin occupations were computed by projecting the electronic
wavefunctions on the atomic orbitals of the iron, as prescribed by equation (3).
The upper panel of Fig.~\ref{symm} represents the occupations of the minority spin
manifold in the quintuplet state. Iron(II) is a d$^6$ ion and the quintuplet
bears four unpaired electrons, therefore the sum of the occupations on
the minority spin channel should be around 1. It is not exactly 1 because the
eigenstates of the complex do not correspond to pure d atomic orbitals, but are 
instead strongly hybridized. Yet, it is possible to assign the electronic configuration
of the system in terms of d atomic orbitals, depending on whether the occupations
are close to 0 or 1. Note that by convention, the nitrogen atoms of the porphyrin ring are
placed on the $xy$ plane, with the $x$ and $y$ axes oriented along the Fe-N bonds.
Pure DFT ($U$=0), and DFT+U with $U<$ 2 eV, converge to the 
(d$_{z^2}$)$^1$ (d$_{xy}$)$^1$ (d$_\pi$)$^3$ (d$_{x^2-y^2}$)$^1$ state.\cite{dpi}
This is the same configuration as reported by
Spiro and coworkers from B3LYP simulations.\cite{spiro}
The increase of $U$ above 2 eV stabilizes the 
(d$_{z^2}$)$^1$ (d$_{xy}$)$^2$ (d$_\pi$)$^2$ (d$_{x^2-y^2}$)$^1$ state,
which is the one experimentally assigned to Fe$\mathbf{^{II}}$P(Im).\cite{revscheidt}
Interestingly enough, this change in configuration is associated with an
inversion in the relative energy of the triplet and the quintuplet, which
now becomes the ground state. 

\begin{figure} \centerline{
\resizebox{4in}{!}{\rotatebox{-90}{\includegraphics{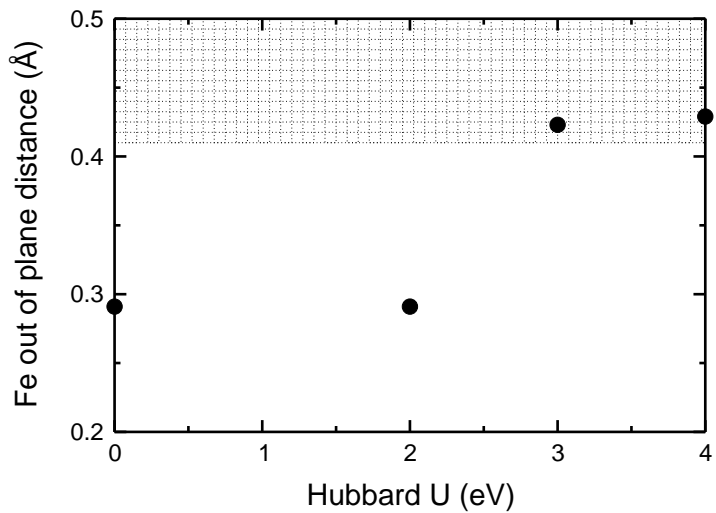}}}}
\caption{Distance of the iron to the average plane
defined by the four nitrogen atoms of the porphyrin ring as a
function of $U$ in Fe$\mathbf{^{II}}$P(Im). The shaded area encompasses the experimental region.}
\label{dfep}
\end{figure}

The examination of the
optimized geometry at a finite $U$ of 3.9 eV shows an agreement with
the experimental data at least as good as pure DFT does.
The structural parameters most affected by the $U$ correction
are those in the vicinity of the metal center, presented in Table III.
The out-of-plane displacement of the iron, d$_{Fe-p}$,
is the distance of the iron to the average plane
defined by the four nitrogen atoms of the porphyrin ring. The interplay between
spin state and d$_{Fe-p}$, often involved in the
dynamics of the heme protein---as for example in the
allosteric mechanism of hemoglobin\cite{strayer}---has been characterized
experimentally\cite{revscheidt} and
theoretically.\cite{rovira, spiro} Table III contrast this and other optimized
structural parameters with the experimental data available for the synthetic model compound
Fe$\mathbf{^{II}}$TPP(2-MeIm) 
(TPP: tetraphenyl porphine; 2-MeIm: 2-methyl imidazole).\cite{FeTPP2MeIm}
Fig.~\ref{dfep} shows the dependence of d$_{Fe-p}$ on $U$, with the shaded part of 
the graph indicating the experimental region.

In summary: the $U$ term favors the stabilization of the (d$_{xy}$)$^2$ 
configuration---deemed the experimental ground state of the 
complex---rendering this (quintuplet) state the lowest in energy, as
can be seen in the lower panel of Fig.~\ref{symm}.

\subsection{Iron(III) porhyrinato chloride}

The low lying accessible electronic states for the penta-coordinate Fe$\mathbf{^{III}}$P(Cl)
complex, with five d-electrons, are the sextet, the quartet, and the doublet
(S=5/2, S=3/2 and S=1/2, respectively). Similarly to the situation discussed in the
previous section, DFT can not reproduce the high spin character of the system, which
has been established experimentally.\cite{exp_fepcl} 
Using a battery of ab-initio methods, Ghosh and collaborators have explored this complex
in depth.\cite{ghosh,ghosh2,ghosh3} They found that, while the PW91 exchange-correlation
functional yields a quartet
state 8.1 kcal/mol more stable than the sextet, B3LYP provides nearly
identical energies for both configurations. Higher-level CASPT2 calculations and 
CCSD(T) simulations on a smaller model system are
consistent with experiments, placing the sextet almost 20 kcal/mol below the
quartet.\cite{ghosh2, ghosh3} These results are summarized in Table IV.

\begin{figure} \centerline{
\resizebox{4in}{!}{\rotatebox{-90}{\includegraphics{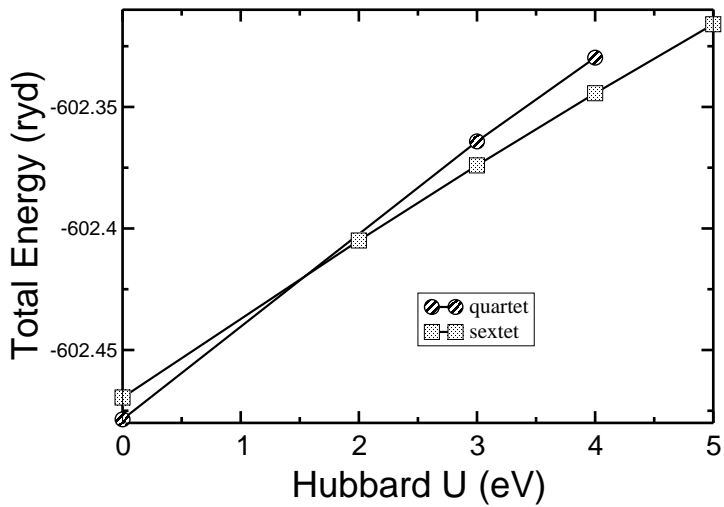}}}}
\caption{Total energy of the lowest accessible spin states of the
Fe$\mathbf{^{II}}$P(Cl) complex as a function of the $U$ parameter.}
\label{fepcl}
\end{figure}

Fig.~\ref{fepcl} shows that, as seen in Fe$\mathbf{^{II}}$P(Im),
the $U$ term stabilizes the highest multiplet in Fe$\mathbf{^{III}}$P(Cl).
A spin inversion is verified at $U \approx$ 1.5 eV, rendering the
sextet as the ground state. A value of $U_{sc}$ equal to 4.0 eV is
obtained, which leads to a sextet-quartet transition energy of 9.2 kcal/mol.
Table IV makes evident the poor performance of density-functionals to describe
multiplet splittings in transition metals,
capable of errors in the order of tens of kcal/mol.
Despite its quantitative disagreement with the highly correlated methods
(whose ultimate accuracy is, on the other hand, difficult to assess in this case),
DFT+U succeeds in recovering the ordering of the spin states.

\subsection{Six coordinated oxyheme model}

The Fe$\mathbf{^{II}}$P(Im)(O$_2$) system has been long identified as low
spin in native proteins and in synthetic compounds.\cite{revscheidt}
Its importance as the oxygenated model of hemoglobin and myoglobin is reflected in
the literature, which---aside from the experimental work---offers many computational
studies addressing the electronic and structural aspects of the 
complex.\cite{rovira,jibio,ourjacs}
The low spin nature of six coordinated iron porphyrins is
in general correctly described by DFT, consequently with its trend to unstabilize
high multiplicity states. In the particular case of Fe$\mathbf{^{II}}$P(Im)(O$_2$),
calculations with different functionals, including B3LYP, indicate a singlet ground state of
open-shell character.\cite{ijqc,rovira}
While the total spin of the molecule is zero, DFT calculations reveal
partial spin densities localized on the d orbitals of Fe and the $\pi^*$ orbitals 
of O$_2$, integrating approximately to +1 and -1, corresponding
to two unpaired electrons of opposite spin.\cite{ijqc,rovira}
This open-shell singlet (o.s.s.) state can be interpreted as the result of an
antiferromagnetic coupling  between Fe$\mathbf{^{II}}$P(Im) (S=2) and O$_2$ (S=1), 
each retaining part of its magnetic character upon binding.

\begin{figure} \centerline{
\resizebox{3in}{!}{\includegraphics{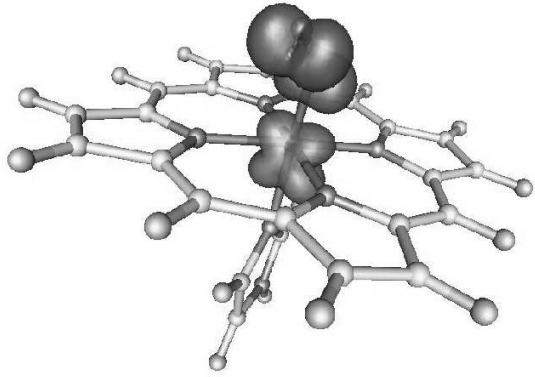}}}
\caption{Spin density in Fe$\mathbf{^{II}}$P(Im)(O$_2$) corresponding to an
open-shell singlet, calculated with DFT+U.
Lobes localized on the iron and on the O$_2$ represent unpaired electron density of
opposite spin.}
\label{spin}
\end{figure}

\begin{figure} \centerline{
\resizebox{4in}{!}{\rotatebox{-90}{\includegraphics{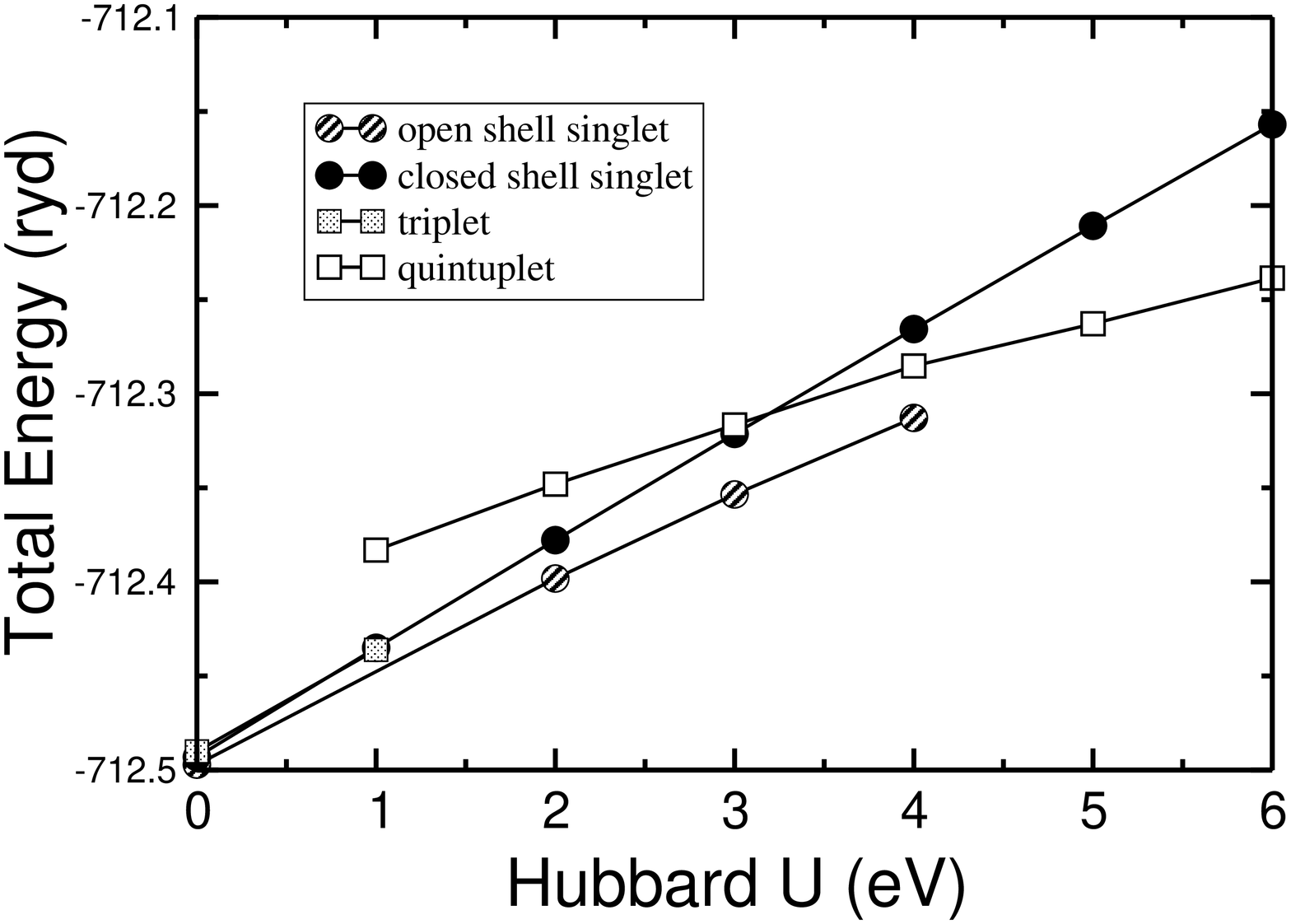}}}}
\caption{Total energy of the low lying spin states of
Fe$\mathbf{^{II}}$P(Im)(O$_2$) as a function of the $U$ parameter.}
\label{oss}
\end{figure}

\begin{figure} \centerline{
\resizebox{4in}{!}{\rotatebox{-90}{\includegraphics{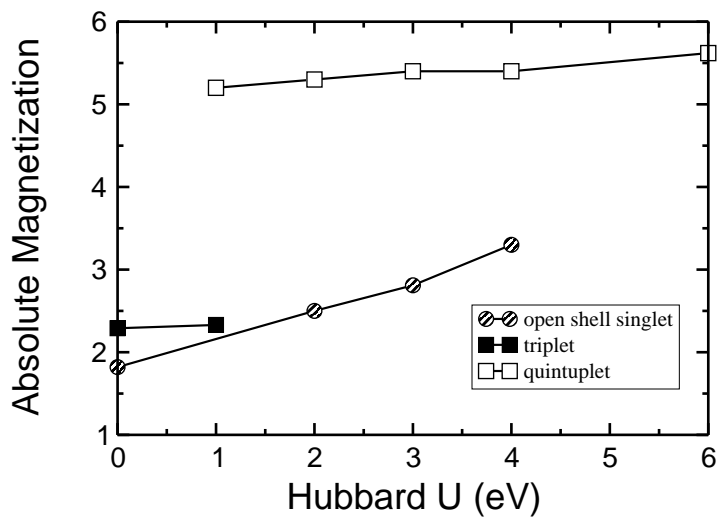}}}}
\caption{Absolute magnetization of the low lying spin states of
Fe$\mathbf{^{II}}$P(Im)(O$_2$) as a function of the $U$ parameter.}
\label{absolute_mag}
\end{figure}

DFT+U supports this picture: Fig.~\ref{spin} depicts the spin density,
$\rho_{spin}({\bf r})=\rho_{\alpha}({\bf r})-\rho_{\beta}({\bf r})$, computed at a finite
$U$ of 4 eV. This figure is qualitatively equivalent to the one reported by Rovira and
coworkers using pure DFT.\cite{rovira}
The impact of the $U$ parameter on the total energies of the lowest accessible spin states
is plotted in Fig.~\ref{oss}. The progressive increment of $U$
further stabilizes the o.s.s. with respect to the closed-shell singlet
and the triplet. On the other hand, the gap between the o.s.s. 
and the quintuplet is reduced, but the raise in $U$ beyond 4 eV produces the 
dissociation of the Fe-O bond in the o.s.s. before a spin
crossing between these two states is observed. 
The effect of the on-site correction can also be examined through the absolute
magnetization of the molecule, defined as 
$\int |\rho_{\alpha}({\bf r})-\rho_{\beta}({\bf r})|d{\bf r}$,
a measure of the unpaired electron density in the system.
Fig.~\ref{absolute_mag} illustrates how the on-site correlation
affects the distribution of the unpaired electron density,
reinforcing, in particular, the antiferromagnetic character of the o.s.s.
The net effect of $U$ is seemingly to thwart the coupling of the unpaired
electrons of the molecular oxygen and the porphyrin, stabilizing the separate
species, which is reflected in the elongation of the Fe-O distance discussed below.
The absolute magnetization augments from 1.8 at $U$=0 to 3.2 at $U$=4 eV, a
value placed halfway from that corresponding to the unbound system, comprised of
a triplet plus a quintuplet (S=1+2).

\begin{figure} \centerline{
\resizebox{4in}{!}{\rotatebox{-90}{\includegraphics{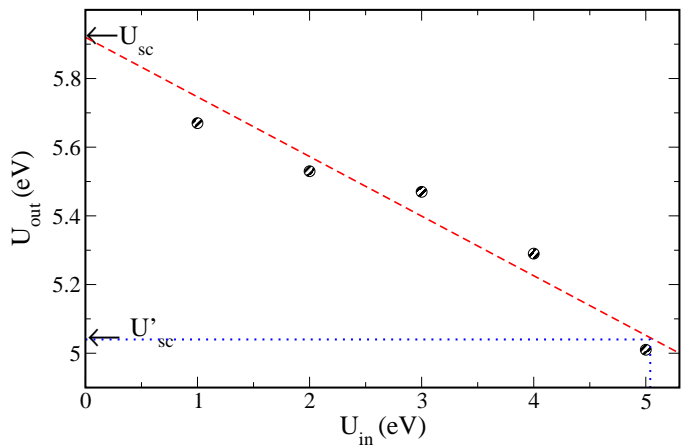}}}}
\caption{Linear response calculation of the $U$ parameter
on the open-shell singlet state of the Fe$\mathbf{^{II}}$P(Im)(O$_2$) complex.}
\label{UvsUo2}
\end{figure}

\begin{figure} \centerline{
\resizebox{4in}{!}{\rotatebox{-90}{\includegraphics{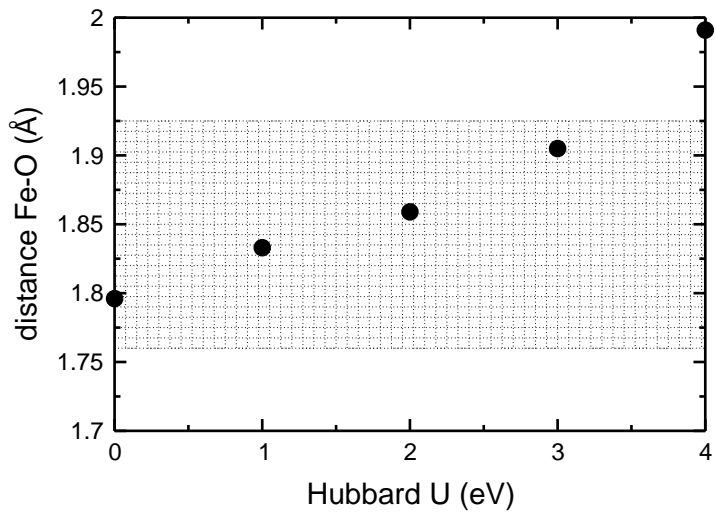}}}}
\caption{Fe-O bond length as a function of the $U$
parameter in Fe$\mathbf{^{II}}$P(Im)(O$_2$).
The shaded area encompasses the experimental region.}
\label{dfeo}
\end{figure}

\begin{figure} \centerline{
\resizebox{4in}{!}{\rotatebox{-90}{\includegraphics{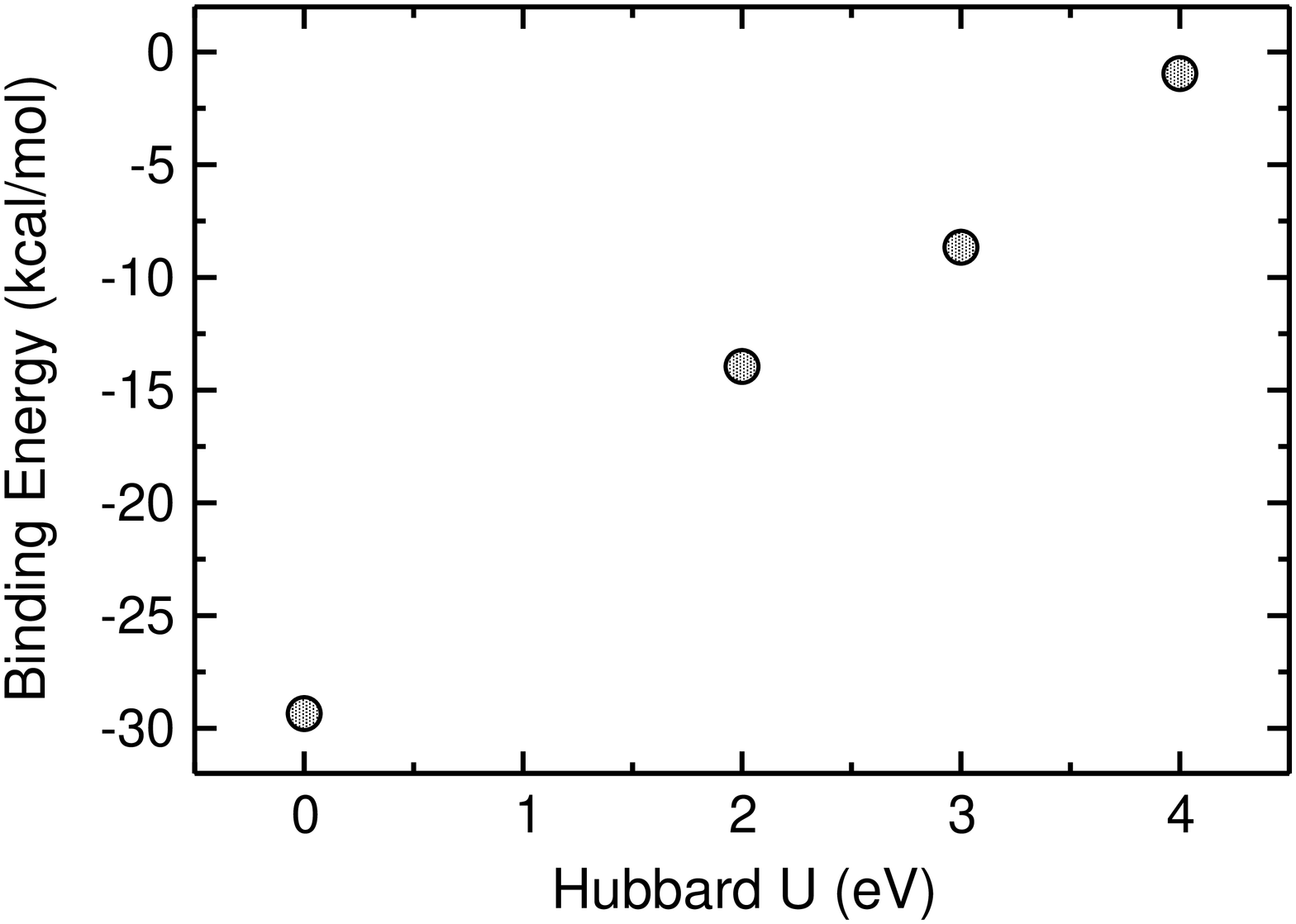}}}}
\caption{Energy of O$_2$ binding to Fe$\mathbf{^{II}}$P(Im) as a function of the $U$
parameter.}
\label{bindenergy}
\end{figure}

Extrapolation to the $y$-axis in the $U_{in}-U_{out}$ plot (Fig.~\ref{UvsUo2})
yields a $U_{sc}$ of 5.9 eV, sensibly higher than in the previous examples.
This value induces the rupture of the Fe-O bond, and in consequence it is
not possible to obtain a relaxed, bound complex associated with this $U_{sc}$.
As seen in Fig.~\ref{dfeo}, where the Fe-O distance is plotted as a function of $U$,
the increase in the on-site correction provokes the elongation of the bond, eventually
leading to the dissociation of the complex. The shaded area in
the graph encloses the range of experimental Fe-O lengths observed in different model
compounds. It is important to emphasize that such values correspond to synthetic or
natural compounds in which the O$_2$ molecule is stabilized in the axial position by
virtue of a second interaction on the distal side, namely an hydrogen bond or some kind
of trapping or cage effect. In the absence of a distal cavity, oxygenation 
of iron (II) porphyrins
under ordinary conditions has been rarely observed.\cite{chemrev94, icbabcock}
Pure DFT and B3LYP systematically overestimate the binding of O$_2$,
giving for the free heme energies in the range of 15-25 kcal/mol,\cite{rovira,jibio}
in direct contradiction with the experimental difficulty to isolate the oxygenated
species. While part of this error
is associated with the underestimation of the total energy of the quintuplet,
such overbinding
represents a major problem in the application of DFT to the calculation
of affinity constants. Fig.~\ref{bindenergy} shows that
DFT+U provides a more realistic oxygen affinity, the binding
energy decreasing from 28 kcal/mol at $U$=0 to around 1 kcal/mol at $U$=4 eV.
In the present case, $U'_{sc}$ turns out to be
1 eV lower than $U_{sc}$ (Fig.~\ref{UvsUo2}). A $U$ of 5.8 eV, as obtained from equation (6),
is too high to represent the thermodynamic and geometrical
properties of the oxygenated complex. This is evincing a positive bias in the
linear response approach, which will be manifested also in other low spin systems.
We will come back to this issue later in the next sections.

\subsection{Five coordinated carboxyheme model}

\begin{figure} \centerline{
\resizebox{4in}{!}{\rotatebox{-90}{\includegraphics{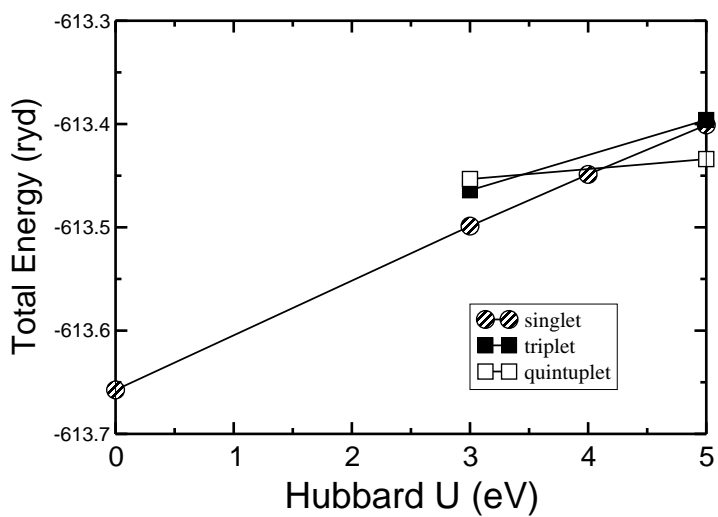}}}}
\caption{Total energy of the low lying spin states of
Fe$\mathbf{^{II}}$P(CO) as a function of the $U$ parameter.}
\label{fepco}
\end{figure}

As the last case study, we will briefly discuss the Fe$\mathbf{^{II}}$P(CO) complex.
Five and six coordinated carboxyhemes are low spin systems whose
electronic and geometrical features are well reproduced by DFT, notwithstanding
the overestimation of the CO binding energy, similarly to what is found with O$_2$.
The motivation to include the carboxylated complex in this study is therefore to
assess the behavior of DFT+U in comparison with standard
density-functional theory, in particular to examine if the former produces any
detrimental bias in a case where the latter shows already a good performance.
Table V contains computed and experimental values for a few selected properties
of the carboxyheme. The general agreement between the simulations and the X-ray data
is benefited from the $U$ term, which not only provides a marginal improvement on the
geometrical parameters of the complex, but also corrects for the overbinding
trend exhibited by pure DFT. At the same time, however, the enhancement of the on-site correlations
closes the gap between the singlet and the quintuplet, to the extent that
for $U \geq$ 4 eV the latter becomes the most stable state (Fig.~\ref{fepco}).

\begin{figure} \centerline{
\resizebox{4in}{!}{\rotatebox{-90}{\includegraphics{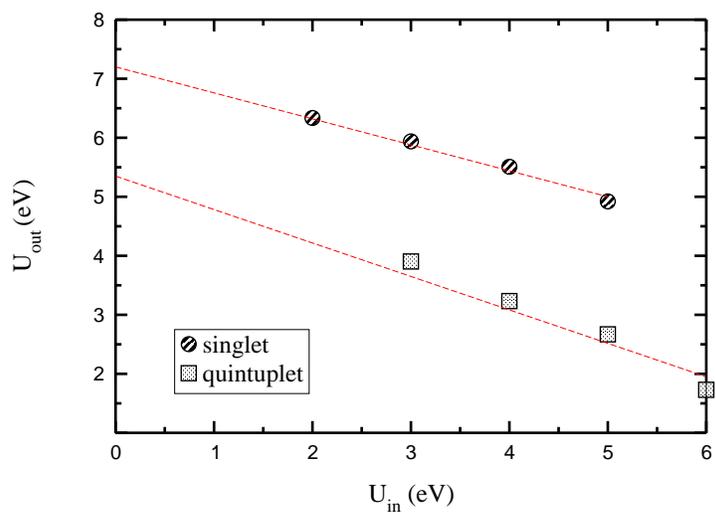}}}}
\caption{Linear response calculation of the $U$ parameter
on the singlet and quintuplet states of the Fe$\mathbf{^{II}}$P(CO) complex.}
\label{UvU-co}
\end{figure}

The response of the system to the change in orbital occupations
resembles the case of Fe$\mathbf{^{II}}$P(Im)(O$_2$), yielding for
$U_{sc}$ and $U'_{sc}$ values of 7.2 and 5.3 eV respectively (Fig.~\ref{UvU-co}).
Since neither experimental, nor reliable theoretical estimations of the spin transition energies
are available, we can not evaluate the magnitude of the error in the self-consistent $U$
obtained with each criteria. In principle, only values below 4 eV
are consistent with the experimental singlet state, and so it is evident that
both $U_{sc}$ and $U'_{sc}$ suffer from some overestimation. Interestingly enough,
the self-consistent $U$ parameter calculated in a high spin configuration of the
carboxyheme turns out
to be significantly smaller, as depicted also in Fig.~\ref{UvU-co}.
On the other hand, linear response calculations on the low spin state of the Fe$\mathbf{^{II}}$P(Im)
system return values of $U_{sc}$ above 6 eV (data not shown). This is indicating
that the response of the system depends more on the multiplicity than on the particular
geometry, coordination mode, or the nature of the ligands.

\section{Conclusions and final remarks}

The inclusion of on-site correlations via a Hubbard term in DFT
rectifies the trend of density-functionals to overstabilize low spin states
in iron porphyrins. 
At variance with DFT, with Hartree-Fock, and with hybrid methods, which successfully describe
some of the possible coordination modes of the complex but fail in the rest,\cite{ijqc}
DFT+U is capable to provide the qualitatively correct splittings in low and high
spin iron porphyrins at the same time, if the proper parameter is adopted. 
This improvement is also reflected in the geometry optimizations,
and, more importantly, in more realistic binding energies to diatomic ligands.
The question of whether a hybrid functional with the proper exchange and correlation
contributions would be capable to recover the spin state energetics of the full iron porphyrins
series is still open. We have addressed this question in a previous
article,\cite{ijqc} with no positive results. In our experience, the combination of
the HF exchange with the GGA exchange-correlation leads to hybrid methods reflecting
either the behavior of pure DFT---overstabilizing low spin states---or the
behavior of Hartree-Fock---favoring high spin states. To the best of our
knowledge, no hybrid has been reported that retains the best of both approaches
in the description of iron porphyrins,
but a more extensive search is probably needed before giving a definite answer.

The application of the linear response calculation to
low spin states leads to self-consistent Hubbard parameters 1 or 2 eV
above the optimal ones. The linear response of the system appears to be more dependent
on the spin state than on the coordination number or the identity of
the ligands. In fact, plots of $U_{in}$ versus $U_{out}$ belonging to different
complexes in the same spin state exhibit a high similarity.
The reason for the overestimation of the Hubbard $U$ in low spin configurations
is not evident. Different extensions to the linear response approach were
explored, including the partition of $U$ into $U^{\alpha}$ and $U^{\beta}$ to discriminate
between both spin channels, and even between the five d states. Additionally,
a $J$ term to represent separately the on-site
exchange was implemented. The modifications described above, however, produced
little or no effect on the resulting $U_{sc}$. The investigation of other factors
which could be responsible for these biases is in progress. In any case,
values of $U$ of 4 eV or slightly lower seem the optimal to reproduce the
electronic, thermodynamic and structural properties of the heme compounds.

\section{Acknowledgments}

This research was partially supported by grants from Fundaci\'{o}n Antorchas and
CONICET. DAS is a member of the scientific staff of CONICET (National Scientific
Council, Argentina).

\newpage

\begin{table}
\caption{Experimental and calculated electronic ground states of
five and six coordinated iron porphines (FeP), with the following axial
ligands: O$_2$, CO, imidazole (Im), and chloride.}
\begin{center}
\begin{ruledtabular}
\begin{tabular}{lccccc}
 & \multicolumn{2}{c} {Six coordinated} & \multicolumn{3}{c} {Five coordinated}\\
\cline{2-3} \cline{4-6}
 & FeP(Im)(O$_2$) & FeP(Im)(CO) & FeP(CO) & FeP(Im) & FeP(Cl) \\
\cline{1-6}
 Experimental & singlet & singlet & singlet & quintuplet & sextet \\
 Hartree-Fock & quintuplet & quintuplet & quintuplet & quintuplet & sextet \\
 DFT-GGA & singlet & singlet & singlet & triplet & quartet \\
 B3LYP & singlet & singlet & singlet & triplet & quartet/sextet \\
\end{tabular}
\end{ruledtabular}
\end{center}
\end{table}

\begin{table}
\caption{Spin transition energies (kcal/mol) for the low lying spin states of
Fe$\mathbf{^{II}}$P(Im) calculated with several density-functionals and with DFT+U, using
$U_{sc}$=3.9 eV.}
\begin{center}
\begin{ruledtabular}
\begin{tabular}{lcccc}
  & & Singlet & Triplet & Quintuplet \\
\hline
    & PBE$^a$ & 7.8 &  0.0  & 7.9   \\
DFT & BP86$^b$ & 8.3  & 0.0  & 6.5  \\
    & B3LYP$^c$ & 5.8 & 0.0  & 1.9   \\
\hline
\multicolumn{2}{c} {DFT+U} & 20.9 & 4.9 & 0.0 \\
\end{tabular}
\end{ruledtabular}
\end{center}
$^a$Pseudopotential calculations with plane wave basis set.
$^b$Pseudopotential calculations with plane wave basis set, from ref. \cite{rovira}.
$^c$Gaussian calculations (VTZ basis), from ref. \cite{spiro}.
\end{table}

\newpage

\begin{table}
\caption{Selected experimental and optimized structural parameters (\AA) in
Fe$\mathbf{^{II}}$P(Im).}
\begin{center}
\begin{ruledtabular}
\begin{tabular}{lccc}
   & d$_{Fe-p}$ & Fe-N$_{porph}$ & Fe-N$_{imi}$ \\
\cline{2-4}
 Experimental$^a$ & 0.42 & 2.09 & 2.16 \\
DFT (PBE) & 0.29 & 2.07 & 2.13 \\
DFT+U ($U_{sc}$=3.9 eV) & 0.43 & 2.11 & 2.19  \\
\end{tabular}
\end{ruledtabular}
\end{center}
$^a$Data for Fe$\mathbf{^{II}}$TPP(2Me-Im) from ref. \cite{FeTPP2MeIm}.
\end{table}

\begin{table}
\caption{Spin transition energies (kcal/mol) for the low lying spin states of
Fe$\mathbf{^{II}}$P(Cl) calculated with highly correlated methods and density functional
theory, including DFT+U ($U_{sc}$=4.0 eV).}
\begin{center}
\begin{ruledtabular}
\begin{tabular}{lcc}
  & Quartet & Sextet \\
\hline
CASPT2$^a$ & 19.6 & 0.0   \\
RCCSD(T)$^b$  & 16.1 & 0.0 \\
DFT-PBE & 0.0 &  5.6  \\
DFT-PW91$^c$ & 0.0  & 8.1 \\
DFT+U & 9.2 & 0.0 \\
\end{tabular}
\end{ruledtabular}
\end{center}
$^a$Ref. \cite{ghosh2}. $^b$Calculations on a simplified model, ref. \cite{ghosh3}.
$^c$Ref. \cite{ghosh2}.
\end{table}

\newpage

\begin{table}
\caption{Selected bond distances (\AA) and CO binding energy (kcal/mol)
for the Fe$\mathbf{^{II}}$P(CO) complex calculated with DFT+U.}
\begin{center}
\begin{ruledtabular}
\begin{tabular}{lcccc}
 U (eV) & Fe-C(CO) & Fe-N$_{porph}$ & C-O & $\Delta$E \\
\hline
0.0 & 1.69 & 1.99 & 1.17 & -45.3 \\
3.0 & 1.71 & 2.00 & 1.16 & -26.1 \\
5.0 & 1.74 & 2.01 & 1.16 & -12.0  \\
\hline
Experimental$^a$ & 1.77 & 2.02 & 1.12 & - \\
\end{tabular}
\end{ruledtabular}
\end{center}
$^a$The experimental model is axially coordinated to pyridine, ref. \cite{ibers}.
\end{table}


\begin{thebibliography}{10}
\bibitem{dpi}
We denote by d$_\pi$ the linear combinations d$_{xz} \pm $d$_{yz}$, 
so d$_\pi$ can be occupied by up to four electrons.
\end{thebibliography}
\end{document}